\documentclass[runningheads]{llncs}
\usepackage[T1]{fontenc}
\usepackage{graphicx,booktabs,amsmath,amssymb,url,array}
\usepackage[hidelinks]{hyperref}
\usepackage{microtype}
\usepackage{marvosym}
\usepackage{multirow}
\graphicspath{{figures/}}

\newcommand{\rx}[1]{\texttt{#1}}

\begin{document}

\title{TT-SAX: A Two-Track Time Series Symbolization for Natural Language Shape Querying}
\titlerunning{TT-SAX}

\author{Zhao Tan \and Xianglong Bao \and Zhipeng Xi \and Ming Jin\textsuperscript{(\Letter)}}
\authorrunning{Z. Tan et al.}
\institute{Griffith University, Australia\\
\email{zhao.tan@griffith.edu.au}, \email{xianglong.bao@griffithuni.edu.au},\\
\email{z.xi@griffith.edu.au}, \email{ming.jin@griffith.edu.au}}

\maketitle

\begin{abstract}
Natural language querying for time series databases aims to let users find what they need in long histories of data by asking in plain language.
Shape queries ask how a series moves. They are the main challenge of this task. Recent work makes them feasible by using symbolization as a bridge between the query and the long history.
However, existing symbolizations were designed for compression, similarity indexing, or classification. What a symbolization should provide for natural language querying has not been systematically studied.
We therefore examine the definitions of six classical symbolizations against this task and identify three limitations.
(1)~Direction is missing from the string,
(2)~direction is measured at too small a scale, and
(3)~the letters are arbitrary.
Three design principles follow.
We follow these principles and make a minimal redesign of SAX called TT-SAX (two-track SAX). The level letters stay. Each segment gains a trend letter. It is measured across neighboring segments rather than inside one, and graded to match the adjectives that queries use.
We conduct extensive experiments on the public NLQTSBench benchmark, comparing TT-SAX with the six classical symbolizations across six LLMs. TT-SAX outperforms all six on shape queries, and ablations trace its gains back to the three limitations one by one.
This paper is the first systematic study of time series symbolization with natural language querying as the design goal. It provides a foundation for future research on this task. Our code is available at \url{https://github.com/Atlamtiz/tt-sax}.
\end{abstract}
\section{Introduction}
\label{sec:intro}

Time series are among the fastest growing kinds of data. Sensors, servers, and markets produce them continuously, and time series databases (TSDBs) now store histories of billions of points.
Querying has not kept pace with storage. ``On which day of the past year was the closing price highest?'' asks about a value, and one ordinary SQL query answers it. On the same curve, users often want to ask about shapes such as ``on which days did the price plunge and then recover quickly?'' Answering such shape queries at database scale remains an open challenge.
This paper is concerned with letting non-expert users pose shape queries to a TSDB in a single sentence. This is the main challenge within natural language querying for time series databases (NLQ4TSDB) \cite{sonarts}.

\begin{figure}[t]
\centering
\includegraphics[width=1\linewidth]{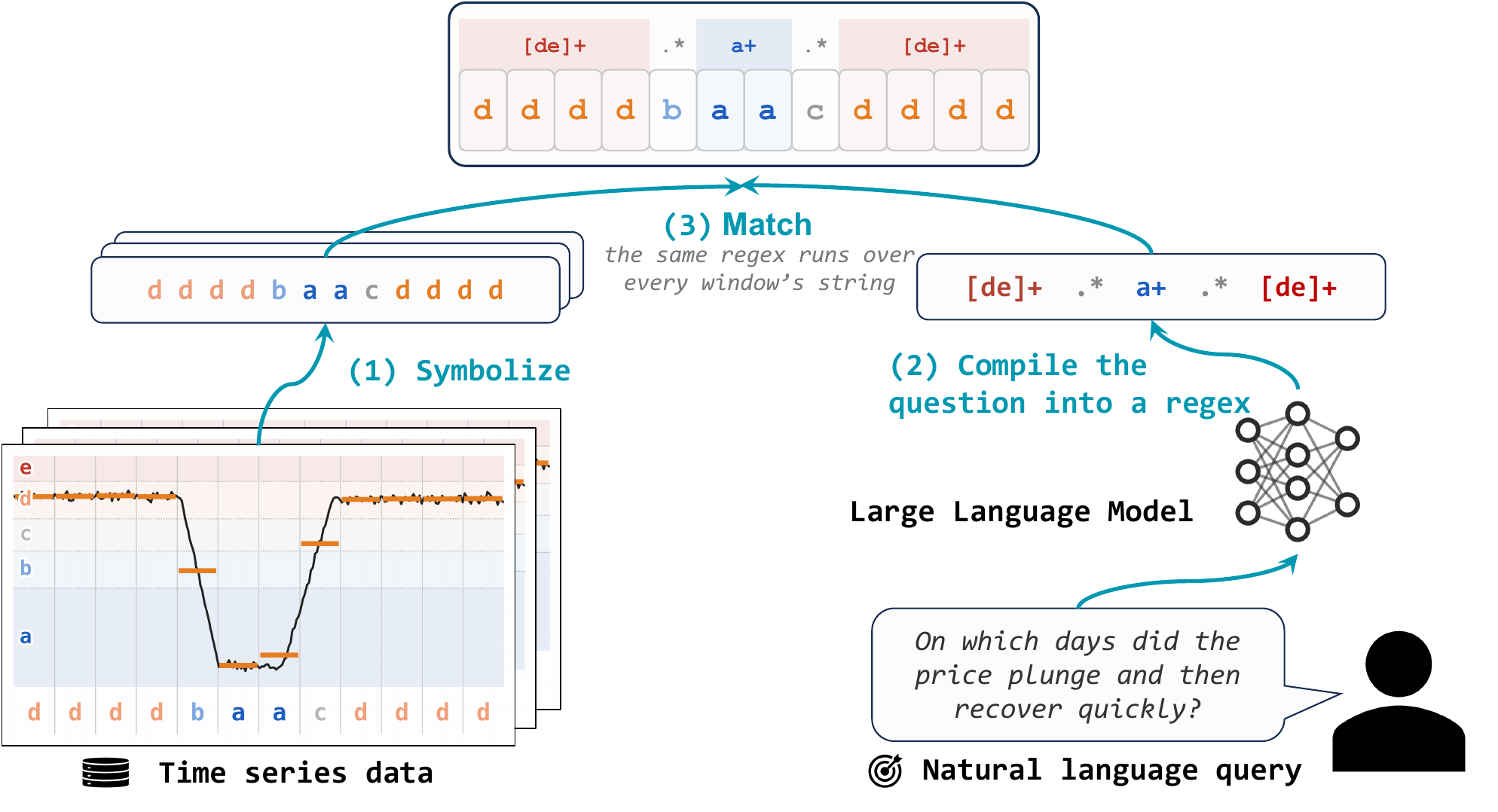}
\caption{How symbolization serves a natural language shape query, illustrated with SAX. (1)~Each window of the series is symbolized into a short string of letters. (2)~An LLM compiles the query into a regular expression over the letters. (3)~The regex runs over every window's string.}
\label{fig:bridge}
\end{figure}

The recently proposed Sonar-TS \cite{sonarts} answers shape queries at database scale by using symbolization as a bridge between language and data.
On one side, the time series is compressed offline into short strings of discrete symbols. On the other, a large language model (LLM) compiles the query into a regular expression over these symbols.
Figure~\ref{fig:bridge} shows an example. The query above, ``plunge and then recover quickly'', compiles to \rx{[de]+.*a+.*[de]+}. This means a run of high letters, then bottom letters, then high letters again.
In this way, a shape query over a long history of raw values becomes one regular expression matched against short strings. This search is cheap enough to cover an entire database.

Sonar-TS adopts the most basic symbolization, SAX, and leaves open which symbolization suits NLQ4TSDB.
Symbolization itself has a long history. SAX \cite{sax} writes one level letter per segment of a window. Its descendants add more numeric detail to the string, such as extrema, slopes, angles, and adaptive pieces \cite{esax,onedsax,tfsax,tsax,abba}. These additions serve distance and classification, never an LLM.
We argue that these symbolizations run into three limitations here.
(1)~\emph{Direction is missing.} In level-only symbolizations such as SAX and ESAX \cite{esax}, a rise never appears as a symbol; it is only implied by the order of level letters. A regex must then guess the level spellings of a rise, and either covers them all and matches too much, or covers a few and misses.
(2)~\emph{Direction is measured at too small a scale.} 1d-SAX \cite{onedsax}, TFSAX \cite{tfsax}, and TSAX \cite{tsax} do write direction, but they measure it inside a single segment. A segment is short, so the measurement follows the noise inside a long trend, and one sustained rise is written as a broken run of trend letters. Longer segments would smooth the measurement, but they would also coarsen the level letters and erase short shapes.
(3)~\emph{The letters are arbitrary.} A shape query ultimately lands on the trend letters, yet most existing symbolizations spell them with abstract codes. For example, 1d-SAX writes a steep rise as the letter ``e''. For distance functions and classifiers this is harmless, since renaming every letter changes nothing. For an LLM, whether the alphabet itself helps or hurts retrieval is a question that has received little study.

The three limitations point to three principles. Direction should be written into the string. It should be measured at the scale that queries describe. And the letters should be chosen for the LLM, not left arbitrary. We follow these principles with a minimal redesign of SAX, called TT-SAX (two-track SAX). Each window is written as two words. The level word is the SAX word, untouched. The trend word gives each segment one trend letter, measured across neighboring segments. The letter has five grades spelled F, f, s, r, R for steep fall, slow fall, stable, slow rise, and steep rise. Trend letters are not new. We choose and grade them for the LLM.

Our contributions are as follows.
\begin{itemize}
\item We analyze six classical symbolizations against NLQ4TSDB and identify three limitations, which yield three design principles.
\item We propose TT-SAX, a minimal redesign of SAX that follows these principles. It pairs the level word with an aligned trend word, measured across neighboring segments and spelled with a mnemonic alphabet.
\item We evaluate TT-SAX against the six classical symbolizations on the public NLQTSBench benchmark across six LLMs, with ablations attributing its gains to the three design choices.
\end{itemize}

\section{Background and Related Work}
\label{sec:related}

\subsubsection{Natural language querying for time series.}
Querying a TSDB in plain language splits into value queries and shape queries. Value queries inherit mature Text-to-SQL technology \cite{nl2sqlsurvey}; shape queries do not, and systems for them are recent. Sonar-TS \cite{sonarts} searches with a regular expression over stored SAX words and verifies the surviving candidates on their raw values. Its planner is guided by \emph{skills}, guideline lists learned automatically in the spirit of Reflexion \cite{reflexion}. The same work introduces NLQTSBench, whose composite trend subset is our workload. Sonar-TS fixes the symbolization to SAX and does not ask whether SAX is the right choice.

\subsubsection{Symbolizations.}
SAX \cite{sax} z-normalizes a window, averages it over $w$ segments, and maps each mean to a letter with Gaussian breakpoints. Its descendants add information per segment, such as minimum and maximum letters (ESAX \cite{esax}), a regression slope letter (1d-SAX \cite{onedsax}), a trend angle letter (TFSAX \cite{tfsax}), or the sign of the slope (TSAX \cite{tsax}). TVA \cite{tva} pairs each SAX letter with a mnemonic U/D/S letter from the slope inside the segment. Zalewski et al. \cite{zalewski} measure across segments. They quantize the differences between successive segment means with data-driven clusters. Their measurement is the closest precedent of our trend word. ABBA \cite{abba} abandons the fixed grid for an adaptive polygonal chain whose pieces are clustered into letters. Frequency and projection based symbolizations such as SFA \cite{sfa} and SPARTAN \cite{spartan} are excluded by construction. Their symbols index Fourier or principal coefficients, so a time-ordered shape cannot be written as a pattern over them. Comparisons of symbolizations \cite{spartan} judge them by reconstruction, distance, and classification. We found no prior work that designs or evaluates a symbolization for use by an LLM.

\subsubsection{Shape queries.}
Shape querying is older than natural language interfaces. Every line of work meets the same two obstacles, scale and authoring. Formal languages such as SDL \cite{sdl} require the user to write the shape as an expression. We borrow the term shape query from SDL. SAXRegEx \cite{saxregex} runs regular expressions over SAX strings and comes closest to our substrate. However, its regex is derived from an example pattern that the analyst supplies and tunes in a visual tool. Sketch interfaces such as Qetch \cite{qetch} ease authoring but not scale. ShapeSearch \cite{shapesearch} accepts natural language through a fixed parser into a closed algebra over in-memory trendlines. Qute \cite{qute} maps trend vocabulary onto numeric shapes. TSseek \cite{tsseek} reaches database scale with a spatial index, but its queries are formal trend expressions, not language. Neural aligners such as time series language models \cite{chatts} and contrastive retrievers \cite{clasp} take language directly. But they run a model on every candidate window, a cost that does not scale to a database. Each line addresses one obstacle at most. The symbolic bridge addresses both. An LLM authors an open-ended pattern from the query. The pattern runs over a symbol index at database scale.

\section{TT-SAX}
\label{sec:ttsax}

\begin{figure}[t]
\centering
\includegraphics[width=.9\linewidth]{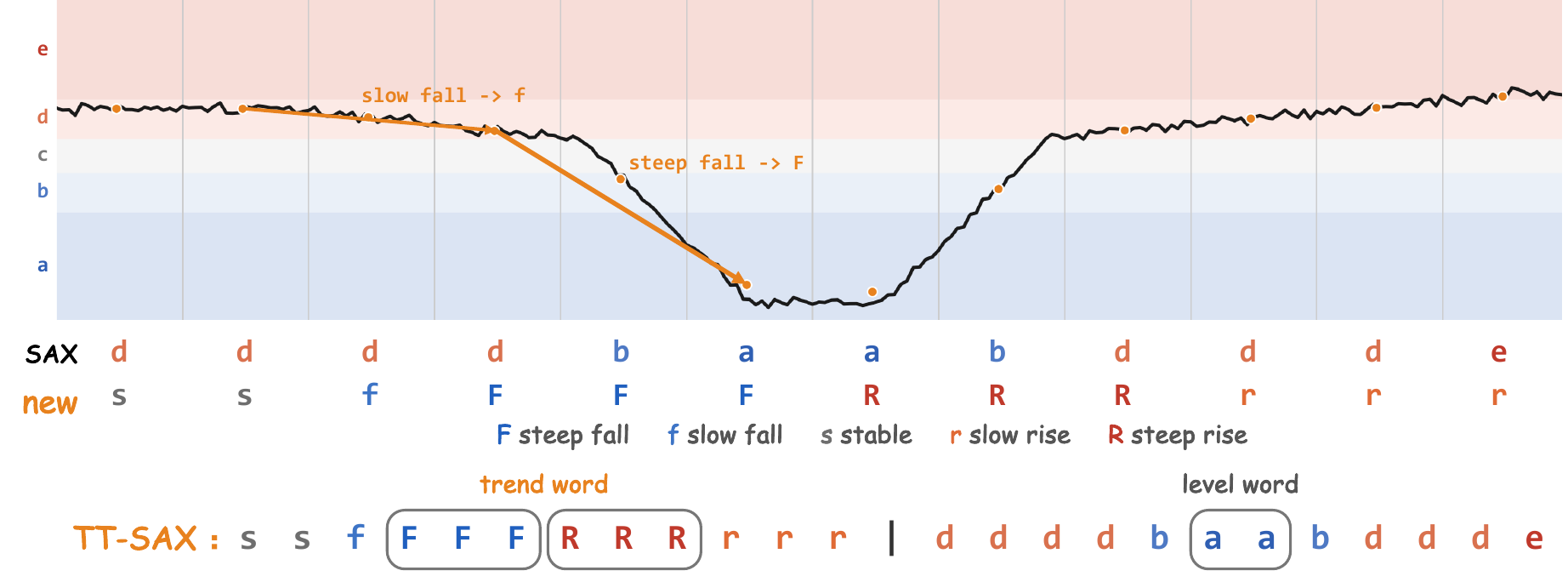}
\caption{An illustration of TT-SAX. Each window is written as two words, a trend word with one trend letter per segment, and the level word.}
\label{fig:ttsax}
\end{figure}

TT-SAX (two-track SAX) writes each window as two words separated by a bar, as Figure~\ref{fig:ttsax} shows. The \emph{level word} is the SAX word itself, $w$ letters from Gaussian breakpoints on the window-normalized segment means. The \emph{trend word} adds one trend letter per segment.

TT-SAX adds the trend word to the symbol string to meet the needs of natural language shape queries. TVA, 1d-SAX, TFSAX, and TSAX also write a trend letter, but they measure the trend inside a single segment.
A segment is short, so these measurements are sensitive to the noise inside a long trend.
TT-SAX instead measures the trend of segment $i$ across its neighbors, as the centered difference $d_i=(p_{i+1}-p_{i-1})/2$ of the real-valued segment means $p_1,\dots,p_w$. At the two ends we use the one-sided differences $d_1=p_2-p_1$ and $d_w=p_w-p_{w-1}$.
The trend letter thus summarizes a two-sided context and stays aligned with the level letter at the same position.
In essence this is smoothing, and it matches the macroscopic way natural language describes a shape. We analyze the number of neighboring segments in Section~\ref{sec:ablation}.

Each $d_i$ then becomes one of five letters. Two fixed thresholds split the size of the change into stable, slow, and steep. They are $\tau_1=0.10$ and $\tau_2=0.30$ on the window-normalized scale. The window is normalized first, so steep means steep relative to that window's own variation, not an absolute physical rate. The sign of the change picks the direction. Both values were fixed before any experiment and were never selected on development or test labels. Section~\ref{sec:widen} reports a post-hoc sensitivity check. The base letters are mnemonic, \rx{f} for fall, \rx{s} for stable, and \rx{r} for rise. Case separates slow from steep, giving \rx{F}, \rx{f}, \rx{s}, \rx{r}, \rx{R}. The alphabet is symmetric between rises and falls and is written for the LLM. The alphabet affects how well an LLM understands and uses the trend letters, and Section~\ref{sec:ablation} analyzes alternative alphabets.

Cross-segment slope information is not new by itself. Zalewski et al.~\cite{zalewski} discretize the first differences between successive segment means, $p_{i+1}-p_i$.
TT-SAX uses a related signal for a different purpose. Its difference is centered, so the direction belongs to the central segment and aligns with its level letter.
Its grades come from fixed, symmetric thresholds, not from learned clusters.
Its letters are chosen for natural language.
The contribution is not a new numerical operation but this combination. A contextual trend measurement whose width is decoupled from the level grid, two-track alignment, and a language-oriented alphabet together serve natural language shape querying.

\section{Experimental Setup}
\label{sec:setup}
\subsubsection{Dataset.}
We use NLQTSBench \cite{sonarts}, the only public benchmark of NLQ4TSDB. Its composite trend subset provides 220 queries over multi-domain real-world channels. Each query names a trend pattern of two to four parts, drawn from \{rise, fall, stable\} with adjectives \{rapid, slow, steady, fluctuating\}. It asks for the days in a long history that show this pattern, for example ``among days with a rapid rise then a stable stretch, the five where the rise is fastest''.
We also synthesize 1{,}000 fresh queries with the benchmark's public generator and reserve them for the stress test in Section~\ref{sec:robust}.

\begin{figure}[t]
\centering
\includegraphics[width=1\linewidth]{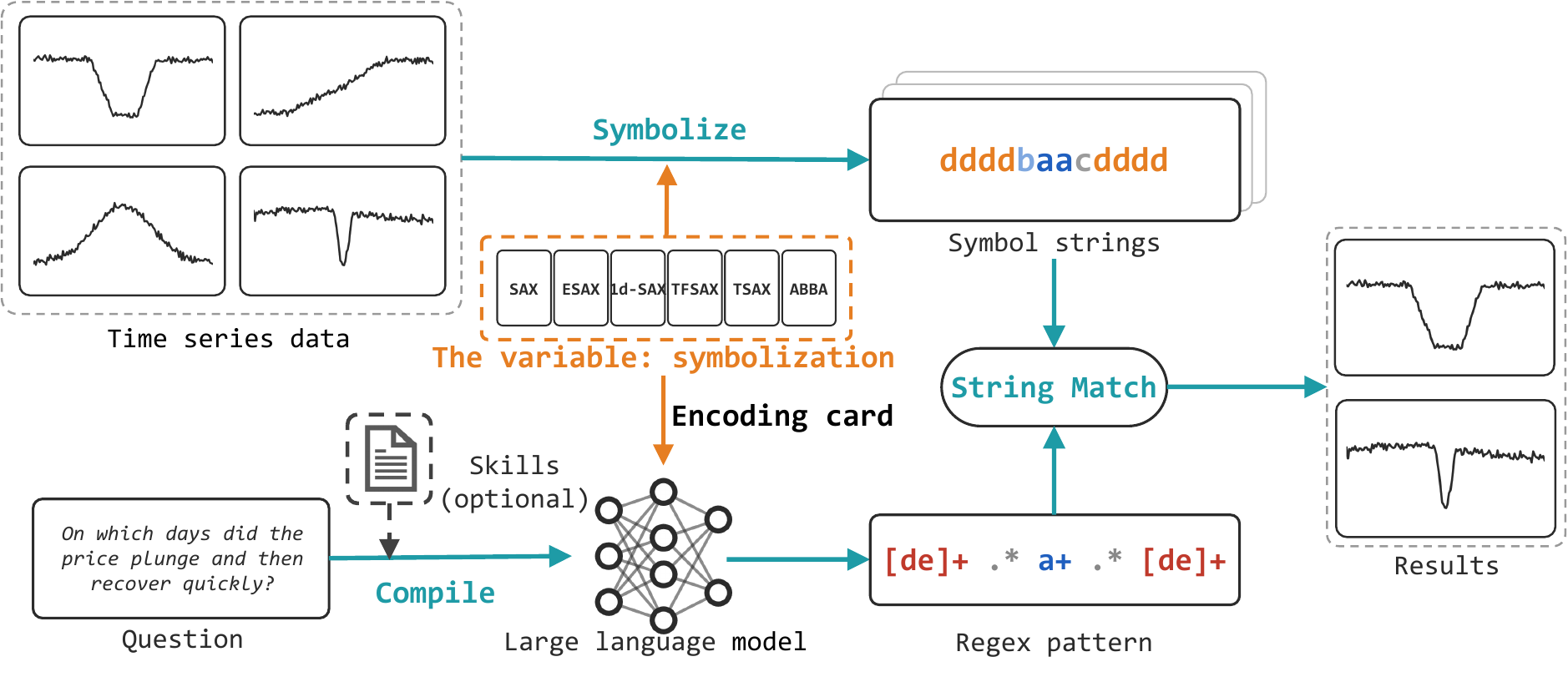}
\caption{The evaluation pipeline. The only variable is the orange slot, the symbolization. It determines both the symbol strings and the encoding card the model reads.}
\label{fig:pipeline}
\end{figure}

\subsubsection{Pipeline.}
Figure~\ref{fig:pipeline} shows the pipeline that every experiment shares.
In the \emph{symbolize} step, every time series is turned into symbol strings. The symbolization that does this is the variable of the study.
In the \emph{compile} step, an LLM turns the query into one regular expression over the letters. Its prompt consists of the query, the \emph{encoding card}, and \emph{skills}.
The encoding card is the short manual of the chosen symbolization. For example, it covers how many levels there are and what each letter means.
Skills are usage guidelines learned beforehand on development queries, separately for each symbolization. Skills are optional, and with them a symbolization comes closer to its ceiling.
In the \emph{match} step, the regular expression runs over every string, and the matching windows are the candidates.
In short, everything in this pipeline is shared except the symbolization. The strings and the card follow from that choice. Any difference in what the search returns therefore traces back to the symbolization.

\subsubsection{Metrics.}
Symbolic search is the first stage of search-then-verify. It returns candidate windows, and a verifier then checks them on the raw values. A missed answer is lost, while an extra candidate only costs verification. A good search therefore keeps all the answers and few other candidates. \emph{Recall} measures the first, \emph{candidates} the second, and \emph{lift} reads the two as one number.
\begin{itemize}
  \item \emph{Recall}: the fraction of a query's answer windows that the search keeps.
  \item \emph{Candidates}: the fraction of all windows that the search keeps.
  \item \emph{Lift}: recall divided by candidates, with a floor on candidates. This is the lift metric of data mining, the precision of the search divided by the precision of a random filter. For a query with recall $R_q$ and candidates $C_q$,
  \begin{equation}
  L_q=\frac{R_q}{\max(C_q,\,10\%)},
  \qquad
  \mathrm{lift}=\frac{1}{n}\sum_{q=1}^{n}L_q ,
  \label{eq:lift}
  \end{equation}
  where $n$ is the number of queries, and the floor of 10\% was fixed in advance to stop a pattern that matches almost nothing from scoring high by luck.
\end{itemize}

\subsubsection{Baselines.}
We compare against six classical symbolizations.
SAX~\cite{sax} is the base. It cuts a day into $w$ segments and writes the mean of each segment as one of $\alpha$ level letters.
ESAX~\cite{esax} writes three level letters per segment, its minimum, mean, and maximum, in the order they occur.
1d-SAX~\cite{onedsax} writes two letters per segment, one for the level and one for the slope of a fitted line.
TFSAX~\cite{tfsax} also writes a second letter, taken from the angle of the segment's net change.
TSAX~\cite{tsax} appends `u' or `d' to each level letter, for whether the segment goes up or down.
ABBA~\cite{abba} drops the fixed grid. It covers the day with a few adaptive linear pieces, clusters them, and writes one letter per piece, so its strings are short and vary in length.
All symbolizations including TT-SAX run at one shared setting with no per-method tuning, namely $w=24$ segments per day where a fixed grid applies and $\alpha=5$ letters per attribute.

\subsubsection{Model.}
The main experiments use DeepSeek-V4-Flash. For robustness across models, Section~\ref{sec:robust} repeats the comparison with DeepSeek-V4-Pro, Gemini-3.7-Flash, GPT-5.6-Luna, Qwen3.6-35B, and Claude Haiku~4.5. All models are called through their APIs in August and September 2026 with temperature zero and reasoning disabled.

\subsubsection{Skills learning.}
The skills in the prompt are learned beforehand, separately for each symbolization, following Sonar-TS's recipe. We generate 80 development queries with the benchmark's generator. The skills are distilled from the model's own failures on them over four rounds. The learning is repeated with three seeds, and we report test means. 

\section{Results}
\label{sec:results}

\subsection{Main Results}
\label{sec:main}
Table~\ref{tab:main} reports the three metrics for TT-SAX and the six baselines, with and without skills.
Without skills, SAX, TSAX, and ABBA are no better than a random filter, whose lift is 1 by definition. ESAX and TFSAX do only slightly better. 1d-SAX reaches the highest recall, but it buys that recall with high candidates. TT-SAX matches its recall with far fewer candidates. Its lift of $2.42$ is $1.7\times$ that of 1d-SAX, and every baseline stays below $1.5$.
With skills every symbolization improves, and the weakest improves most. ABBA rises from $0.37$ to $0.69$ recall. ESAX keeps the fewest candidates, but its recall is now the lowest. 1d-SAX again pays for its recall with candidates. TT-SAX reaches the best recall while its candidates barely grow. Its lift of $3.04$ is $1.4\times$ that of the next best.

Two conclusions follow.
First, doing well on one metric alone is easy. A looser pattern matches more windows, so recall rises but candidates rise with it. This is how 1d-SAX and TSAX score. A tighter pattern shrinks candidates but loses answers too. This is how ABBA and ESAX score. The hard part is doing well on both at once. TT-SAX is the only symbolization that does, so its lift stands apart.
Second, skills change the numbers and reorder several baselines. ABBA rises from last to fourth on lift and ESAX falls to last, yet TT-SAX keeps the highest lift under both conditions. Skills and symbolization interact, but skills do not close the gap. On lift, TT-SAX outperforms every other symbolization on natural language shape queries.

\begin{table}[t]
\centering
\caption{Main results of TT-SAX. Best in bold, second best underlined. $n=220$ queries; every metric is averaged over queries; the results with skills average three skill-learning seeds, the results without skills are single runs.}
\label{tab:main}
\setlength{\tabcolsep}{6pt}
\begin{tabular}{@{}l ccc @{\hspace{16pt}} ccc@{}}
\toprule
 & \multicolumn{3}{c}{w/o skills} & \multicolumn{3}{c}{w/ skills} \\
\cmidrule(lr){2-4}\cmidrule(lr){5-7}
Symbolization & Recall $\uparrow$ & Cand. $\downarrow$ & Lift $\uparrow$ & Recall $\uparrow$ & Cand. $\downarrow$ & Lift $\uparrow$ \\
\midrule
SAX    & 0.43 & 35\% & 1.00 & 0.59 & 36\% & 1.56 \\
ESAX   & 0.47 & 34\% & 1.19 & 0.49 & \textbf{28\%} & 1.53 \\
1d-SAX & \textbf{0.73} & 51\% & \underline{1.42} & \underline{0.83} & 49\% & 2.11 \\
TFSAX  & 0.55 & 38\% & 1.39 & 0.59 & \underline{33\%} & \underline{2.16} \\
TSAX   & 0.62 & 50\% & 1.01 & 0.65 & 43\% & 1.78 \\
ABBA   & 0.37 & \textbf{25\%} & 0.94 & 0.69 & 41\% & 1.94 \\
\midrule
TT-SAX (ours) & \underline{0.71} & \underline{32\%} & \textbf{2.42} & \textbf{0.87} & 36\% & \textbf{3.04} \\
\bottomrule
\end{tabular}
\end{table}

\subsection{Ablation Studies}
\label{sec:ablation}
We ablate TT-SAX in three ways. We remove one of the two words, widen the trend measurement, and change the trend letters. All ablations run without skills, which removes the learning loop as a source of variation.

\subsubsection{Removing a track.}
Table~\ref{tab:tracks} reports the ablation of the two words.
Removing the trend word hurts most. Without it the model has to spell the trend through sequences of level letters. These patterns keep many windows yet catch few answers. Lift falls to $1.00$, the level of a random filter.
Removing the level word also lowers recall, but the candidates shrink with it. The lift holds at $2.03$, above every baseline without skills in Table~\ref{tab:main}. This means the trend word alone still finds answers efficiently, which supports the premise of TT-SAX. What natural language shape querying needs most is a direct expression of the trend.

\begin{table}[ht]
\centering
\caption{Ablation of the two words.}
\label{tab:tracks}
\setlength{\tabcolsep}{8.5pt}
\begin{tabular}{@{}l ccc@{}}
\toprule
Variant & Recall $\uparrow$ & Cand. $\downarrow$ & Lift $\uparrow$ \\
\midrule
TT-SAX & 0.71 & 32\% & 2.42 \\
\midrule
w/o trend word & 0.43 & 35\% & 1.00 \\
w/o level word & 0.55 & 26\% & 2.03 \\
\bottomrule
\end{tabular}
\end{table}

\subsubsection{Widening the trend measurement.}
\label{sec:widen}
\begin{figure}[t]
\centering
\includegraphics[width=.8\linewidth]{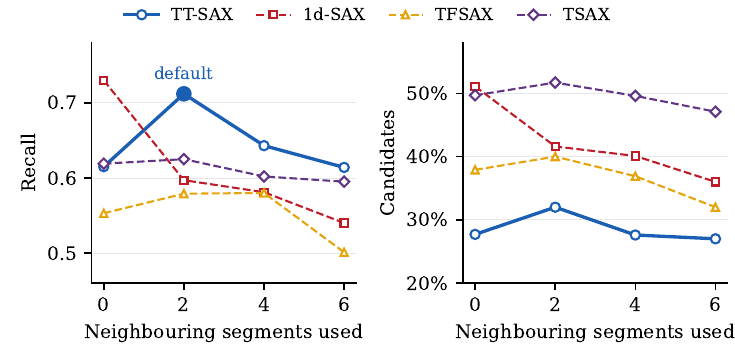}
\caption{Recall (left) and candidates (right) as the trend measurement widens, for TT-SAX and the three baselines that also write a trend letter.}
\label{fig:scale}
\end{figure}

TT-SAX measures the trend of a segment across its neighbors rather than inside the segment alone. The choice follows from how queries are worded. Natural language describes macroscopic movement, and a segment is short, so the change inside it is closer to noise than to trend.
We test the choice by widening the trend measurement of TT-SAX and of the three baselines that also write a trend letter. Each uses its own formula, refitting its slope, angle, or sign over the wider span. Figure~\ref{fig:scale} reports the results.

TT-SAX's recall rises and then falls as neighboring segments are added, and is best at two neighbors. This supports the choice behind the trend word. Inside a single segment, noise dominates the change.
We also note that the best neighbor count is not universal. It depends on the query distribution, the data, and the other parameters of the symbolization. We therefore keep the shared settings of the NLQTSBench benchmark throughout and do not search this space.
The baselines also write trend letters, but they define them differently, so widening does not give them a comparable gain.
TSAX is the clearest case. Its trend alphabet has two letters, up and down. A wider span is usually a slower one, but the alphabet has no letter for slow. Every widened span is still written as up or down, and nothing is gained.
1d-SAX grades its slope on a scale computed for the segment's length, so the widened slopes no longer fit the scale. TFSAX divides its angle by the turning points the wider span brings in.
These results motivate decoupling the trend measurement from the level grid.

The trend word has one more parameter, the two thresholds that cut the size of the change. Five other pairs, from $0.05/0.20$ to $0.15/0.40$, give recall between $0.65$ and $0.68$, candidates between 31\% and 33\%, and lift between $2.11$ and $2.30$. Each of these lifts is above every baseline without skills. The default pair was fixed before the experiments and was left unchanged.

\subsubsection{Changing the letters.}
In natural language shape querying, every symbolization comes with an encoding card. The card tells the model what each letter means, and the model reads it to compile the query into a regex.
Whether the choice of letters changes this compilation is unknown.
Classical symbolization never faced this question. Its letters go into distance computations, where a letter is just a name. We test thirty alphabets in three groups, one per design philosophy.
Group~I gives the letters meaning. Each is the first letter of a direction word, as in TT-SAX's \rx{FfsrR}, with \rx{f} for fall and \rx{r} for rise.
Group~II drops the meaning but keeps the pairing. The letters still form a fall family and a rise family by case, as in \rx{KkxmM}, yet they start no word.
Group~III drops the pairing too. The letters are a plain alphabet run, as in \rx{vwxyz}, and only their order separates fall from rise.
Eight case-paired alphabets are additionally rewritten with their cases swapped.

\begin{figure}[t]
\centering
\includegraphics[width=1\linewidth]{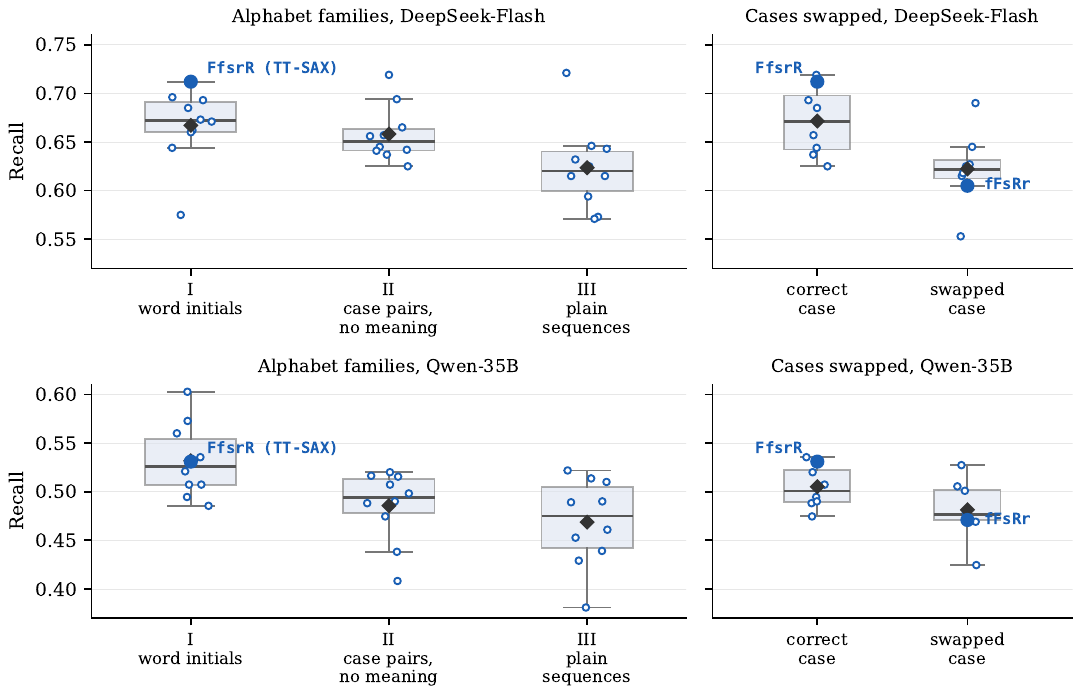}
\caption{Ablation of the trend letters. Group~I letters are the initials of direction words, e.g., \rx{FfsrR}. Group~II letters pair by case but carry no meaning, e.g., \rx{KkxmM}. Group~III letters are plain alphabet runs, e.g., \rx{vwxyz}. In the case-swapped alphabets the uppercase and lowercase of every case-paired letter are exchanged, so \rx{FfsrR} becomes \rx{fFsRr}.}
\label{fig:letters}
\end{figure}

We run the thirty alphabets, ten per group, on one fixed symbolization and two LLMs, DeepSeek-V4-Flash and Qwen3.6-35B. Figure~\ref{fig:letters} reports the results.
First, Group~I is the best group with both LLMs, and its lead grows on the weaker one. A plausible reading is that a weaker model follows the card less reliably, so letters that already carry the meaning help it most.
Second, the assignment of uppercase and lowercase matters. Rewriting TT-SAX's \rx{FfsrR} as \rx{fFsRr} drops recall from $0.71$ to $0.61$. The average loss across the eight swapped pairs is five points. The two alphabets carry the same information, and each encoding card states its mapping precisely. The loss therefore shows that the LLM does not take the mapping from the card alone. It is consistent with a preferred reading of the two cases, which the swapped alphabets go against.
In summary, renaming the letters has no effect on distance and compression, the classical uses of symbolization. For natural language shape querying it does. The TT-SAX alphabet is built for this. Its direction words give the LLM the meaning it needs, its uppercase letters keep the convention the LLM expects, and the whole alphabet stays readable to the human.

\subsection{Skills Transfer}
\label{sec:transfer}

\begin{figure}[t]
\centering
\includegraphics[width=1\linewidth]{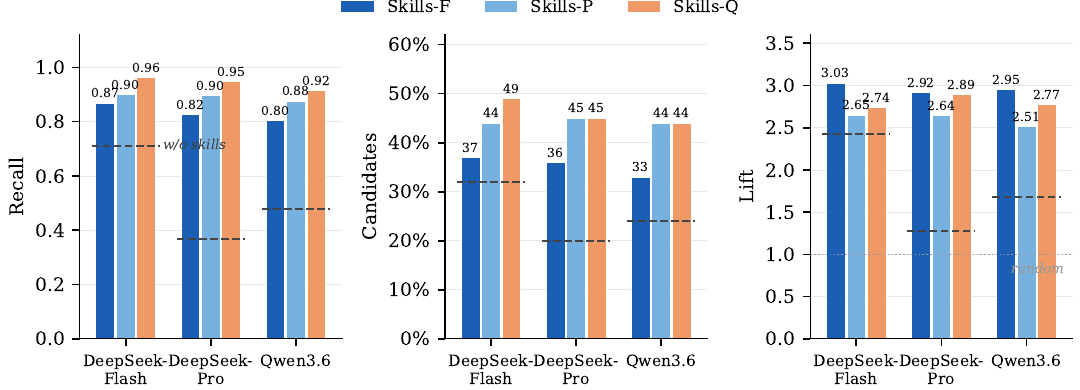}
\caption{Skills transfer results of TT-SAX.}
\label{fig:transfer}
\end{figure}

An LLM compiles the query into a regex by reading the encoding card and the skills. The skills are learned by self-evolution on development queries with Sonar-TS's recipe~\cite{sonarts}. The model answers, analyzes its failures, and summarizes skills round after round. Models differ in ability and style, so the skills they evolve differ too. Can the skills learned by one model serve another? Three models learn their own TT-SAX skills on the same development queries. DeepSeek-V4-Flash, DeepSeek-V4-Pro, and Qwen3.6-35B give Skills-F, Skills-P, and Skills-Q. Each model then uses all three sets. Figure~\ref{fig:transfer} reports the results.

Three findings follow. First, the skills are transferable. Every combination of model and skills beats that model without skills. Second, a set of skills keeps its character on every model. With Skills-Q every model reaches its highest recall, and with Skills-F every model keeps its fewest candidates. Third, a model's own skills are not necessarily its best. On lift, Skills-F is the best set for all three models. For DeepSeek-V4-Pro and Qwen3.6-35B it beats their own skills.

\subsection{Robustness Analysis}
\label{sec:robust}

\begin{figure}[t]
\centering
\includegraphics[width=.9\linewidth]{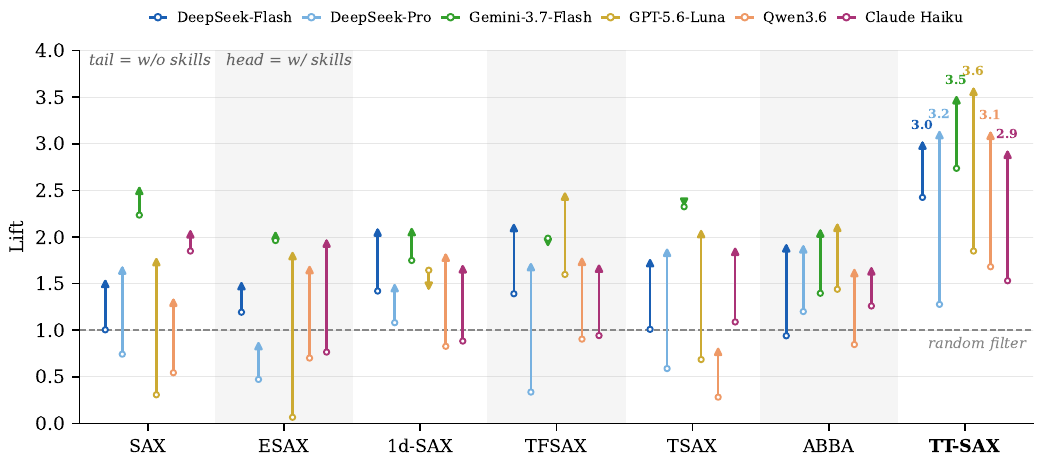}
\caption{Lift of every symbolization under six LLMs, without skills (arrow tail) and with skills (arrow head).}
\label{fig:models}
\end{figure}

We assess TT-SAX under five further LLMs. We replay both conditions with prompts identical to the main experiment. Figure~\ref{fig:models} reports the results. TT-SAX is robust. Its average lift over the five models is $1.81$ without skills and $3.28$ with, against $1.24$ and $1.93$ for the best classical symbolization.

\begin{figure}[t]
\centering
\includegraphics[width=.7\linewidth]{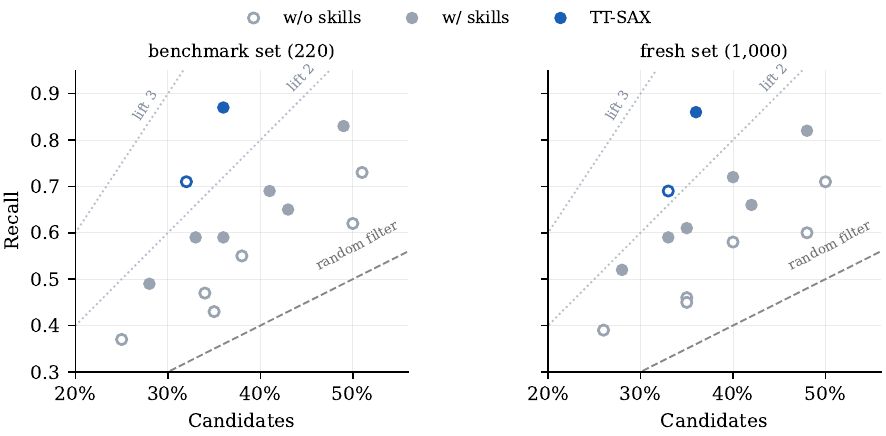}
\caption{Stress test on a fresh set of 1{,}000 queries.}
\label{fig:stress}
\end{figure}

Most of our experiments run on the public benchmark, and its 220 queries are few. A symbolization could look strong by fitting the particulars of a small query set, and a larger set could give a different result. We therefore use the reserved set of 1{,}000 queries from the benchmark's public generator~\cite{sonarts}. We deduplicate them against the benchmark and development queries, so the set is larger but not differently distributed. We rerun TT-SAX and all six baselines on it. Figure~\ref{fig:stress} shows the results. Every symbolization moves by at most three points in recall and in candidates. The skills help as before. TT-SAX keeps its place nearest the upper-left corner. It is the only symbolization with lift above two on both sets in both conditions. The advantage holds on a query set nearly five times larger.

\section{Conclusion}
\label{sec:conclusion}
This paper studied how time series symbolization should be designed for natural language shape querying. We systematically analyzed the limitations of classical symbolizations in this setting. We then proposed TT-SAX, which represents level and contextual trend information in two aligned tracks. Further analyses indicate that its gains are associated with an explicit trend word, an appropriate measurement span, and meaningful letters. The gains remain consistent across six LLMs. The results demonstrate the importance of designing symbolizations that are both numerically informative and accessible to LLMs.

\begin{credits}
\subsubsection{\ackname} This research was supported by the NVIDIA Academic Grant in Higher Education and Developer Program and the CSIRO--National Science Foundation (US) AI Research Collaboration Program.
\end{credits}

\bibliographystyle{splncs04}
\bibliography{refs}
\end{document}